\RequirePackage{fix-cm}
\documentclass[twocolumn,epjc3]{svjour3}  
\RequirePackage[T1]{fontenc}
\smartqed  
\usepackage[utf8]{inputenc}
\RequirePackage{graphicx}
\usepackage{subcaption}
\RequirePackage{mathptmx}      
\RequirePackage{flushend}
\RequirePackage[colorlinks,citecolor=blue,urlcolor=blue,linkcolor=blue]{hyperref}
\usepackage{microtype} 
\usepackage{siunitx}
\usepackage[modulo]{lineno}
\usepackage{textcomp}
\usepackage[dvipsnames]{xcolor}
\usepackage{booktabs}
\usepackage{makecell}
\usepackage{amsmath}
\usepackage{amssymb}
\newcommand{\ReD}{\mbox{ReD}}
\newcommand{\LArTPC}{\mbox{LAr~TPC}}
\newcommand{\TPC}{\mbox{TPC}}

\newcommand{\SiPMs}{\mbox{SiPMs}}
\newcommand{\Am}{$^{241}$Am}

\newcommand{\Er}{$E_r$}
\newcommand{\Erbar}{$\bar{E}_r$}
\newcommand{\Cf}{$^{252}$Cf}
\newcommand{\Qy}{$Q_{y}$}
\newcommand{\Nel}{$N_e$}

\newcommand{\fprompt}{$f_{p}$}
\newcommand{\tdrift}{$T_\mathrm{drift}$}

\newcommand{\xy}{$x-y$}

\newcommand{\BaF}{BaF$_{2}$}
\newcommand{\gtwo}{$g_2$}

\journalname{Eur. Phys. J. C}
\makeatletter

\begin{document}

\title{Characterization of the ionization response of argon to nuclear recoils at the keV scale with the ReD experiment}




\author{P.~Agnes\thanksref{addgssi}
  \and
 I.~Ahmad\thanksref{addastrocent}
 \and
 S.~Albergo\thanksref{addrunict,addrinfnct}
  \and
 I.~Albuquerque\thanksref{addusp}
  \and
 M.~Atzori~Corona\thanksref{addca}
 \and
 M.~Ave\thanksref{addusp,addgssi}
  \and
 B.~Bottino\thanksref{addinfnge,addunige}
 \and
 M.~Cadeddu\thanksref{addca}
 \and
 A.~Caminata\thanksref{addinfnge}
 \and
 N.~Canci\thanksref{addinfnna}
 \and 
 M.~Caravati\thanksref{addgssi}
 \and
 L.~Consiglio\thanksref{addlngs}
 \and
 S.~Davini\thanksref{addinfnge}
 \and
 L.K.S.~Dias\thanksref{addusp}
 \and
 G.~Dolganov\thanksref{addgrigory,addgrigory2}
 \and
 G.~Fiorillo\thanksref{addunina,addinfnna}
 \and
 D.~Franco\thanksref{addapc}
 \and
 M.~Gulino\thanksref{addkore,addlns}
 \and
 T.~Hessel\thanksref{addapc,addceep}
 \and
 N.~Kemmerich\thanksref{addusp}
\and
M.~Kimura\thanksref{addastrocent}
\and
M.~Ku\' zniak\thanksref{addastrocent}
\and
M.~La~Commara\thanksref{adduninafarmacia,addinfnna}
\and
J.~Machts\thanksref{addapc}
\and
 G.~Matteucci\thanksref{addunina,addinfnna}
\and
E.~Moura~Santos\thanksref{addusp}
\and
E.~Nikoloudaki\thanksref{addapc}
\and
 V.~Oleynikov\thanksref{addbudker,addnovo}
 \and
 L.~Pandola\thanksref{addlns}
 \and
 R.~Perez~Varona\thanksref{addusp}
 \and
 N. Pino\thanksref{addrunict,addrinfnct,addlns}
 \and
 S.M.R.~Puglia\thanksref{addrunict,addrinfnct}
 \and
 M.~Rescigno\thanksref{addinfnroma}
 \and
 B.~Sales~Costa\thanksref{addusp}
 \and
 S.~Sanfilippo\thanksref{addlns}
 \and
 A.~Sung\thanksref{addprinceton}
 \and
 C.~Sunny\thanksref{addastrocent}
 \and 
 Y.~Suvorov\thanksref{addunina,addinfnna}
 \and
 R.~Tartaglia\thanksref{addlngs}
 \and
 G.~Testera\thanksref{addinfnge}
 \and
 A.~Tricomi\thanksref{addrunict,addrinfnct,addcsfnsm}
 \and
 M.~Wada\thanksref{addastrocent}
 \and
 Y.~Wang\thanksref{addihep,addihep2}
 \and
 R.~Wojaczy\' nski\thanksref{addastrocent}
 \and
 P.~Zakhary\thanksref{addastrocent,addrunict,addrinfnct}
}

\authorrunning{\ReD\ Working Group}
\institute{Gran Sasso Science Institute, L'Aquila AQ 67100, Italy\label{addgssi}
	\and
           AstroCeNT, Nicolaus Copernicus Astronomical Center of the Polish Academy of Sciences, 00-614 Warsaw, Poland\label{addastrocent}
           \and
           Physics and Astronomy Department, Universit\`a degli Studi di Catania, Catania 90123, Italy\label{addrunict}
           \and
           Istituto Nazionale di Fisica Nucleare, Sezione di Catania, Catania 90123, Italy\label{addrinfnct}
           \and
           Instituto de F\'{\i}sica, Universidade de S\~{a}o Paulo, S\~{a}o Paulo 05508-090, Brasil\label{addusp}
           \and
           Istituto Nazionale di Fisica Nucleare, Sezione di Cagliari, Cagliari 09042, Italy\label{addca}
           \and
           Istituto Nazionale di Fisica Nucleare, Sezione di Genova, Genova 16146, Italy\label{addinfnge}
           \and
            Physics Department, Universit\`a degli Studi di Genova, Genova 16146, Italy\label{addunige}
           \and
           Istituto Nazionale di Fisica Nucleare, Sezione di Napoli, Napoli 80126, Italy\label{addinfnna}
           \and     
           INFN Laboratori Nazionali del Gran Sasso, Assergi (AQ) 67010, Italy\label{addlngs}  
           \and
           National Research Centre Kurchatov Institute, Moscow 123182, Russia\label{addgrigory}
 \and
           National Research Nuclear University MEPhI, Moscow 115409, Russia\label{addgrigory2}
 \and           
           Physics Department, Universit\`a degli Studi Federico II, Napoli 80126, Italy\label{addunina}
 \and      
         APC, Université de Paris Cité, CNRS, Astroparticule et Cosmologie, Paris F-75013, France\label{addapc}
           \and
           Universit\`a di Enna KORE, Enna 94100, Italy\label{addkore}
           \and
           Istituto Nazionale Fisica Nucleare, Laboratori Nazionali del Sud, 95123 Catania, Italy\label{addlns}          
           \and
           Mines Paris, PSL University, Centre for Energy Environment Processes (CEEP), 77300 Fontainebleau, France\label{addceep}
            \and
           Department of Pharmacy, Universit\`a degli Studi Federico II, Napoli 80131, Italy \label{adduninafarmacia}          
           \and
           Budker Institute of Nuclear Physics, Novosibirsk 630090, Russia\label{addbudker}
           \and
           Novosibirsk State University, Novosibirsk 630090, Russia\label{addnovo}
           \and  
          Istituto Nazionale di Fisica Nucleare, Sezione di Roma, Roma 00185, Italy\label{addinfnroma}
           \and
           Physics Department, Princeton University, Princeton, New Jersey 08544, USA \label{addprinceton}
           \and           
           Centro Siciliano di Fisica Nucleare e Struttura della Materia (CSFNSM), Catania 90123, Italy \label{addcsfnsm}
           \and           
           Insititute of High Energy Physics, Beijing 100049, China \label{addihep}
           \and
           University of Chinese Academy of Sciences, Beijing 100049, China \label{addihep2}   
}


\date{Received: date / Accepted: date}

\maketitle

\begin{abstract}
  In the recent years, argon-based experiments looking for Dark Matter in the Universe have explored the
  non-standard scenario in which Dark Matter is made by low-mass Weakly Interacting Massive Particles,
  of mass in the range of 1-10~GeV instead of the canonical hundreds of GeV.
  Detecting such particles is challenging, as their expected signatures are nuclear recoils with energies below 10~keV, observable solely
  via ionization. This necessitates a precise understanding of the detector response in this energy regime, which
  remains incomplete for argon.
  To address this, the ReD experiment was developed within the framework of the DarkSide-20k Collaboration to produce and characterize
  few-keV nuclear recoils.
  A compact dual-phase argon Time Projection Chamber (TPC) was irradiated with neutrons from a \Cf\ source, to produce Ar recoils in the energy range of interest via (n,n') elastic
  scattering. A downstream spectrometer composed of 18 plastic scintillators detected the neutrons scattered off Ar nuclei, enabling recoil energy
  reconstruction via two-body kinematics.
  The ionization yield \Qy\ of argon, defined as the number of electrons produced per unit energy deposit, was measured in a model-independent
  way between 2 and 10~keV. 
  These measurements extend direct experimental coverage well below the previous limit of approximately 7~keV. The results are consistent with existing data above 7~keV, while they indicate a higher \Qy\ at lower energies. 
\end{abstract}

\section{Introduction}\label{intro}
One of the most compelling and theoretically well-motivated candidates for Dark Matter is the Weakly Interacting
Massive Particle (WIMP), a non-Standard Model particle with a mass typically ranging from tens of GeV to a few
TeV~\cite{Bertone:2016nfn,Roszkowski:2017nbc,Arcadi:2024ukq}. This range gives rise to the so-called ``WIMP miracle'',
wherein a weak-scale interaction cross-section naturally yields the observed Dark Matter abundance. WIMP searches are
ongoing through various experimental strategies~\cite{Schumann:2019eaa,Cebrian:2022brv,Billard:2021uyg}, notably via dual-phase
Time
Projection Chambers filled with noble liquids such as argon and
xenon~\cite{DarkSide:2018kuk,Aalseth:2017fik,LZ:2024zvo,XENON:2024wpa,PandaX:2024qfu}.\, 
In these detectors, a thin gas layer is formed above the main liquid volume, allowing nuclear recoils from elastic WIMP interactions to be detected through two channels: scintillation and ionization.
The recoil energy produces argon excimers and electron-ion pairs:  excimers yield
a prompt scintillation signal (S1),
while ionization electrons, drifted by a uniform electric field, 
are extracted into the gas phase and accelerated,
producing secondary electroluminescence (S2)~\cite{Buzulutskov:2020xhd}. The S1–S2 time delay, namely the 
electron drift time, encodes the vertical ($z$) position
and the S2 light pattern determines the \xy\ position, enabling full 3D event reconstruction.

The DarkSide-50 experiment~\cite{DarkSide:2018kuk,DarkSide:2014llq} at the INFN Laboratori Nazionali del Gran
Sasso (INFN LNGS) utilized a dual-phase liquid argon TPC (\LArTPC) to search for WIMPs in
the canonical 100~GeV–1~TeV mass range. Although designed and optimized for this regime, the technology demonstrated
strong potential
for extending the search to WIMPs with masses of a few GeV, commonly referred to as ``low-mass WIMPs''. Interest in low-mass WIMPs
has recently grown following increasingly stringent constraints on standard WIMP models and the resulting pressure to
explore new regions of the parameter space. Their experimental detection is particularly challenging, as their expected
nuclear recoil energies (\Er) in argon are below a few keV, thus making scintillation signals nearly undetectable.
The DarkSide-50 Collaboration performed a dedicated search for low-mass WIMPs 
relying exclusively on ionization signals: this made it possible to lower the energy threshold from 
approximately 20~keV to a few hundreds of eV and to achieve the highest sensitivity to date 
for WIMP interactions in the 1.8–3.0~GeV mass range~\cite{Agnes:2018ves,PhysRevD.107.063001}. 
A similar approach is also being pursued in xenon targets~\cite{XENON:2024znc,PandaX:2022xqx}. 

These low-threshold searches hinge critically on a detailed understanding of the ionization yield (\Qy) of nuclear recoils,
defined as the number of electrons produced per unit of recoil energy. Such an understanding is only partially
available for argon, as direct measurements currently only exist down to $\sim 7$~keV from experiments employing
small-scale LAr TPCs: Joshi et al.~\cite{Joshi:2014fna}, ARIS~\cite{Agnes:2018mvl} and SCENE~\cite{Cao:2015ks}.
The DarkSide-50 low-mass WIMP analysis instead relied on an \emph{ad hoc} phenomenological model, based on the Thomas–Imel box
formalism~\cite{Thomas:1987ek}, to extrapolate the \Qy\ of Ar down to approximately 0.5~keV~\cite{PhysRevD.104.082005}. This model
was constrained through Monte Carlo simulations, dedicated \textit{in situ} AmBe and AmC neutron calibrations and the existing
direct measurements mentioned above. Its predictions are sensitive to the
choice of nuclear stopping power model, such as those by Ziegler~\cite{Ziegler:2010bzy}, Molière~\cite{Moliere1947} and
Lenz-Jensen~\cite{Lenz1932,Jensen1932}. 
While consistent with current data, these models yield notably different \Qy\ predictions below 5~keV, a region not yet
directly probed in argon.
A direct measurement of \Qy\ in this low-energy range is of high importance, particularly in view of the upcoming
multi-tonne argon experiment DarkSide-20k~\cite{Aalseth:2017fik,DarkSide-20k:2024yfq} which is being constructed at INFN LNGS
and aims to explore the low-mass WIMP frontier with improved sensitivity. This scientific motivation led to the
Recoil Directionality (ReD) measurement campaign presented in this work, as part of the program of the DarkSide-20k Collaboration.
The ReD experiment operated a small \LArTPC\ at the INFN Sezione di Catania, Italy, in 2023 and collected Ar nuclear recoil data in
this previously unexplored energy range. The TPC used for this project was fully characterized~\cite{Agnes:2021zyq} and subsequently employed in a measurement campaign to constrain a possible directional sensitivity~\cite{DarkSide-20k:2023nla}.

The structure of this paper is as follows: Sects.~\ref{sec:DetectorLayout} and \ref{sec:Detectors} present the conceptual layout of the
measurement and the description of the individual components of the experimental setup.
Sect.~\ref{sec:DataAndResult} outlines the event reconstruction and the selection of the sample of signal events.
Sect.~\ref{sec:MC} details the Monte Carlo framework and its validation. The statistical
analysis and the \Qy\ measurements are presented in Sect.~\ref{sec:StatisticalAnalysis}, followed by discussion and
conclusions in Sect.~\ref{sec:discussion}.


\section{Conceptual design} \label{sec:DetectorLayout}
The primary objective of this experiment is to measure the ionization yield of Ar nuclear recoils (NRs) in the energy range of
2--10~keV. Argon NRs of known energy can be generated through elastic scattering of neutrons, by irradiating the ReD
dual-phase \LArTPC\ with a suitably configured neutron source. The experimental setup and the choice of the neutron source are
therefore optimized to produce argon NRs within the target energy range, as detailed in the conceptual design outlined below.

Neutrons of appropriate energy can be obtained from a radioactive \Cf\ source, which has a half-life of 2.65~years and undergoes 96.9\% $\alpha$-decay and 3.1\% spontaneous fission (SF)~\cite{Firestone1996ToI}. Each SF event releases, on average, 3.76 neutrons, whose energy spectrum follows a Maxwellian distribution from 0 to 13 MeV, with a mean of 2.3 MeV and a most probable energy around 1 MeV~\cite{Mannhart1987}. The accompanying $\gamma$-rays from SF enable efficient event tagging, representing a significant advantage of \Cf\ over conventional AmBe neutron sources.

The neutrons emitted from the \Cf\ source are collimated by a polyethylene shield and directed toward the TPC, located approximately 1~m away.
The produced neutron beam has an opening angle sufficiently wide to illuminate the full active volume of the \LArTPC, thus maximizing the signal rate. As the beam
traverses the TPC, neutrons can undergo elastic scattering (n,n') off argon nuclei, producing nuclear recoils. The scattered
neutrons continue downstream and are detected by a neutron spectrometer positioned roughly 1~m beyond the TPC.

The nuclear recoil energy \Er\ is reconstructed event-by-event via two-body kinematics as:
\begin{equation}
E_{r} = 2 K_{n}\frac{m_n m_{Ar}}{(m_n + m_{Ar})^2} (1 - \cos\theta_{S}),
\label{eq:recoil_energy}
\end{equation}
where $K_{n}$ is the neutron kinetic energy, $\theta_{S}$ is the scattering angle, and $m_n$, $m_{Ar}$ are the neutron and argon nucleus masses, respectively.
The neutron kinetic energy is obtained from the relativistic relation:
\begin{equation}
K_n = (\gamma - 1) m_n c^2,
\label{eq:neutronke}
\end{equation}
where the Lorentz factor $\gamma$ is calculated from the neutron velocity, determined by the measured time of flight (ToF) along the known path
length $D$ between the \Cf\ source and the spectrometer. 
The relevant ToF values range from a few tens to approximately 200~ns.

A set of tagger detectors placed near the \Cf\ source identifies $\gamma$-rays accompanying neutron emission from SF, providing the START signal
for the ToF measurement. The STOP signal is given by neutron detectors which constitute the neutron spectrometer. Both taggers and spectrometer
detectors are fast scintillators, thus achieving sub-ns time resolution to ensure precise velocity reconstruction.

The scattering angle $\theta_{S}$ is determined by the fixed geometry of the spectrometer. 
To target recoil energies of a few keV from incident neutrons with $\sim 1$~MeV kinetic energy, $\theta_{S}$ must be set to approximately $\ang{15}$, as dictated by
Eq.~\ref{eq:recoil_energy}. The spectrometer must be positioned outside the direct line of sight of the collimated beam, to avoid the large background due to unscattered (direct) neutrons.

\section{Experimental setup} \label{sec:Detectors}
\begin{figure*} %
\centering
 \includegraphics[width=0.8\textwidth]{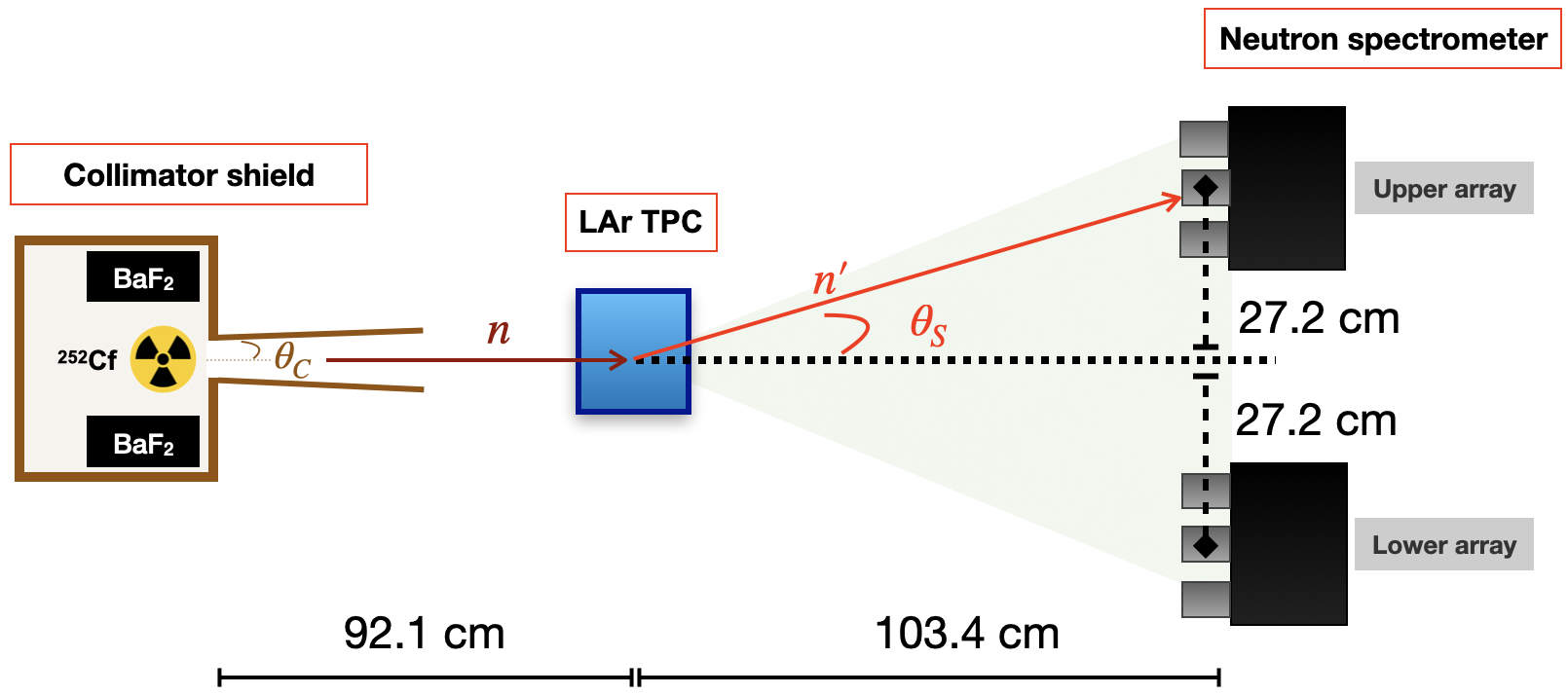}
\caption{Schematic view (not in scale) of the experimental setup. 
On the left is the shielding structure housing the \Cf\ source and the two \BaF\ scintillator taggers. The collimator opening angle $\theta_{C}$ of $\ang{2.6}$
allows the full illumination of the \LArTPC\ (depicted in blue) at 92.1~cm .
On the right, a schematic representation of the neutron spectrometer, not showing the support frame. The two detector arrays are 
mounted symmetrically at $\pm 27.2$~cm above and below the TPC center, outside the cone of the direct neutron flux. 
The spectrometer covers a scattering angle range of $\theta_{S} = \ang{12} - \ang{17}$.  The blue arrow indicates the path of an incoming neutron
from the source prior to (n,n') interaction with Ar in the TPC, while the red arrow depicts a possible trajectory of the scattered neutron
n' within the acceptance of the neutron spectrometer. Cryogenic infrastructure is not shown for simplicity.}
\label{setup}
\end{figure*}

This section describes in detail the components used in the concrete implementation of the conceptual layout of the experiment 
at INFN Sezione di Catania.  
The experimental setup, which is schematically represented in Fig.~\ref{setup}, consists of three main components: the shielding which includes the \Cf\ source with the tagger detectors (left), the \LArTPC\ (middle) and the neutron spectrometer (right).

The \Cf\ source used in this work has an activity of approximately 800~kBq at the beginning of the data taking, 
corresponding to a SF activity of $\sim$25~kBq and about $10^5$~n/s.
It is enclosed within a cubic shielding structure with 15~cm thick walls of boron-loaded high-density polyethylene (HDPE), additionally coated with lead and iron.
Neutrons are largely absorbed by the shielding, except for those collimated into an exit cone ($\theta_C = \ang{2.6}$) carved into a $\sim$50~cm HDPE channel. This configuration ensures uniform irradiation of the ReD TPC, whose center is located 92.1~cm from the source.
Two barium fluoride scintillators (\BaF) placed inside the HDPE shielding, about 3~cm far from the source,
are the taggers that detect the prompt $\gamma$-rays from SF and give the START for the ToF measurement. 
The neutron spectrometer is located at 195.5~cm from the source (103.4~cm downstream of the \LArTPC) 
to provide the STOP signal for the ToF measurement. It is made of 18 plastic scintillators (PScis) 
arranged in two $3 \times 3$ identical arrays mounted in 3D-printed plastic holders. The two arrays are symmetrically placed 
27.2~cm above and below the neutron beam axis within a custom-built light support frame made of aluminum beams.
The detectors span an angular range $\theta_{S}$ from \ang{12} to \ang{17}. They are all positioned outside the direct neutron beam from the collimator, so they are exposed solely to neutrons scattered from the TPC or other components of the setup.
%

The alignment procedure is performed using two laser levels
to ensure that the source (at the collimator exit), 
the center of the \LArTPC\, and the center of the frame of the neutron spectrometer lie on the same 
horizontal optical axis, which coincides with the axis of the conical collimator. 
The alignment between the source and the neutron spectrometer is achieved with a precision of a few mm. 
The alignment of the TPC cannot be performed with comparable accuracy, as the detector is not visible inside the 
cryostat and it is subject to the thermal shrinkage when the system is brought to LAr temperature. The effect 
of a possible mis-alignment of the TPC is accounted for in the data analysis. 



\subsection{The LAr TPC} \label{subsec:TPC}
The core of the experimental setup is the ReD cubic dual-phase \LArTPC~\cite{Agnes:2021zyq}. It has an active volume of 
$5 \times 5 \times 6$~cm$^3$, delimited on the top and bottom by two transparent windows. The windows are coated 
with a thin Tin-Indium-Oxide (ITO) conductive layer, to be operated as anode and cathode, respectively. 

The scintillation and electroluminescence signals are detected by two $5 \times 5$~cm$^2$ tiles of  
Silicon PhotoMultipliers (\SiPMs), placed above and below the active volume, behind the transparent windows. 
Each tile contains 24 custom-made cryogenic \SiPMs~\cite{Gola:2019idb} arranged in a $4 \times 6$ array. The \SiPMs\ of 
the top tile are read out individually in order to enhance the \xy\ spatial resolution, while those of the bottom tile are summed in groups of six and read out as four quadrants.
As the scintillation and electroluminescence emission from Ar takes place 
in the VUV, at 128~nm, all internal surfaces are coated with a layer of Tetraphenyl Butadiene (TPB), which shifts 
the wavelength to about \SI{420}{\nano\meter} to match the sensitivity range of the photosensors.

A layer of gaseous argon of about 7~mm is formed in the upper part of the TPC, above the liquid phase. 
An electrically-grounded stainless steel grid is placed 3~mm below the liquid-gas interface, to separate the drift region 
below from the extraction region above. 

Three electric field regions are established within the \TPC: a drift field 
in the LAr active volume of 200~V/cm; an extraction field of 3.8~kV/cm; and an electroluminescence 
field in the gas phase of 5.7~kV/cm. The maximum drift time at 200~V/cm is 
$T_{\mathrm{driftmax}} $ = 54~\textmu s; this is the time required for an electron produced at the
cathode to reach the liquid surface. The extraction field guarantees 100\% extraction efficiency 
of the electrons from the liquid to the gas phase~\cite{Chepel:2012sj}.

Further details about the ReD TPC and its cryogenic system are provided in
Refs.~\cite{Agnes:2021zyq,DarkSide-20k:2023nla}.

\subsection{The taggers and the neutron spectrometer}
Two \BaF\ inorganic scintillator crystals (BaF0 and BaF1), each with a volume of about 30~cm$^3$,
are placed close to the \Cf\ to act as SF taggers. \BaF\ is chosen for its high $\gamma$-ray efficiency and for its 
fast scintillation properties, which enable operation under high source rates and provide accurate timing. 
The scintillation light has a prominent fast component peaking at \SI{195}{\nano\meter} and sub-ns decay time, and a 
slow component at approximately \SI{310}{\nano\meter}, with a decay time of about \SI{630}{\nano\second}~\cite{baf2}.  
The \BaF\ crystals were wrapped with reflector and optically coupled to Photomultipliers  (PMTs) of the Hamamatsu H11934 series \cite{HamamatsuH11934}. 
For BaF0, a UV-sensitive H11934-100 PMT was employed to enhance  detection of the fast scintillation component and improve timing performance. Since only one PMT of this type was available, BaF1 was coupled to a conventional visible-sensitive H11934-20 PMT.
The sides of the crystals not directly facing the \Cf\ source were shielded using a 2~mm Pb foil, to suppress the background of delayed $\gamma$-rays from radiative 
capture (n,$\gamma$) in the boron of the HDPE shielding.
 
The neutron spectrometer used to detect neutrons scattered off the TPC consists of 18 cylindrical 
EJ-276~\cite{PSci_RYABEVA2021165495} detectors manufactured by Scionix, each coupled to a 
Hamamatsu R1924A PMT~\cite{HamamatsuR1924A}. The organic plastic scintillator EJ-276 is selected for its excellent n/$\gamma$ 
discrimination capabilities by Pulse Shape Discrimination (PSD), which are crucial for suppressing the accidental
$\gamma$-ray 
background. The scintillators have a diameter and length of 1~inch, providing enhanced spatial granularity to determine the neutron position, an essential requirement for accurate kinematic reconstruction.  
The detectors are mounted in two identical $3 \times 3$ arrays, with a 4.3~cm relative spacing in both the horizontal and the vertical directions. 

\subsection{Trigger and data acquisition}
The data acquisition (DAQ) system employs three CAEN V1730 Flash ADC boards (FADC), each providing 16 input channels with 14-bit 
resolution and \SI{2}{\volt} peak-to-peak dynamic range~\cite{CAEN}. 
The three boards are synchronized using an external custom clock module developed by TRIUMF. 
The acquisition software is a customized version of the DAQ code developed by the PADME experiment~\cite{Chiodini_2017}.
Upon a trigger signal, waveforms from all 48~detectors - the two \BaF\ detectors, 28 TPC SiPMs (24 on the top tile 
and 4 on the bottom), and 18 PScis - are digitized at 500~MHz sampling frequency for 40{,}000 samples and stored on 
disk. The acquisition window corresponds to a total duration of \SI{80}{\micro\second}, including a \SI{5}{\micro\second} 
pre-trigger window used for baseline restoration, which is sufficient to fully include events which take place close 
to the TPC cathode and exhibit an electron drift time \tdrift\ close to $T_{\mathrm{driftmax}}$.

The events of interest -- low-energy NRs in the TPC -- are expected to produce a very small S1 signal (about 8 PE for a 5~keV NR), making them difficult to trigger on.
Similarly, it is not efficient to trigger on the delayed S2 signal, which is more conspicuous but distributed 
in a broader time interval. For this reason, the trigger logic is set on the coincidence between either one of 
the two \BaF\ detectors and any of the PSci detectors 
within a $[-56, 200]$~ns time window. 
The gate window is set keeping into account the expected neutron ToF
and the possible occurrence of timing offsets. 
The master trigger signal is generated by using an external standard NIM logic module and 
distributed to the three FADC boards. 
The DAQ thresholds are set above the baseline noise level and regularly adjusted to account for potential 
baseline shifts. The typical energy thresholds are about 100~keV (electron equivalent).
TPC data are acquired in follower mode for each trigger signal, allowing for more efficient offline event building and facilitating the identification of weak TPC signals of interest.

\section{Data taking and selection} \label{sec:DataAndResult}
Data with the \Cf\  source were acquired continuously for about three months, from January to March 2023. 
The analysis in the following is based on 142 individual \Cf\ runs, each lasting about 12~hours, for a total acquisition time of approximately 75~days.
The stability of the SiPM response in the TPC was monitored through weekly calibrations using a 403-nm laser delivered to the inner detector via optical fibers. Similarly, weekly calibrations with a $^{241}$Am source emitting monoenergetic $\gamma$-rays of approximately 60~keV were performed to verify the TPC energy scale.

After data acquisition, candidate NR events in the TPC were identified offline. Firstly, candidate neutron events (``tagged neutrons'') were selected by analyzing the signals from the \BaF\ and PSci detectors, as described in Sect. \ref{subsect:tagged} below. For the population of tagged neutrons, potential NR signals in the TPC  -- S1, S2, or both -- were searched for using the procedures in Sect.~\ref{subsect:tpcreco}. 
Candidate signal events featuring a scattering in the TPC were then selected as described in Sect.~\ref{subsect:finalcuts}, and their recoil energies were reconstructed according to the method in Sect.~\ref{subsect:erec} for the subsequent statistical analysis. 
\subsection{Identification of tagged neutron events} \label{subsect:tagged}
The identification of the candidate neutron events is based on the signals coming from the \BaF\
taggers and the neutron spectrometer. The events of interest are those in which the \BaF\ detects a $\gamma$-ray from the 
\Cf\ SF\footnote{According to Monte Carlo (MC) simulations, about 50\% of \Cf\ SF events produce a signal above 
100~keV in at least one of the \BaF\ detectors} while the PSci detects a neutron. They can be selected by requesting
that the time elapsed between the two signals is consistent with the ToF
of a few-MeV neutron, and that the PSci signal is classified as proton-induced by the PSD. 

\subsubsection{Reconstruction of \BaF\ and PSci signals}
The digitized waveforms from the \BaF\ and PSci detectors were processed by calculating the total charge, integrated within a gate of 2000~ns and 500~ns, respectively.
Energy calibration of the \BaF\ detectors was carried out at the beginning of the experimental campaign using $^{60}$Co, $^{137}$Cs, and $^{152}$Eu $\gamma$ sources. Similarly, the PSci detectors were calibrated with $^{241}$Am and $^{137}$Cs $\gamma$ sources.
The PSci response to proton recoils was measured in a dedicated 
calibration run, in which a detector was directly exposed to neutrons from \Cf; the results were found to be in good 
agreement with the literature values for EJ-276~\cite{PSci_RYABEVA2021165495,Laplace:2020tdj,NGO2023168216}, once the different total 
integration time is accounted for. Based on this calibration, the trigger threshold of 100~keV corresponds to about 
650~keV proton recoils.

The fraction of total charge in the first 30~ns (\fprompt) was used as the PSD discrimination parameter for the
PScis. The short gate was optimized based on the data, resulting in a n/$\gamma$ discrimination better than 2(4)~$\sigma$ above 
200(400)~keV.

The start time of the PSci signals was determined as the interpolated zero-crossing time of the pulse obtained by passing the waveforms through a digital constant fraction discriminator (CFD). The \BaF\ signals are significantly longer (a few \textmu s vs.$~150$~ns) and are produced at a 
10~kHz rate due to the proximity of the \Cf\ source. To determine their start time accurately, a pulse finder based on a moving average algorithm was used to locate the prompt component, which was then passed through the CDF to find the interpolated zero-crossing time.
The detector BaF1 exhibits a weak prompt component in the light
signal, due to the UV-insensitive PMT or possibly to crystal damage. This significantly degrades its timing accuracy compared to the other detector, see Sect.~\ref{subsubsec:timing}. 

\subsubsection{Timing and time resolution} \label{subsubsec:timing}
The synchronization of the FADCs and the time alignment of all readout channels is critical to achieve a resolution at the 1~ns 
level for the ToF measurements. The time alignment of the individual channels was performed by using 
the population of $\gamma$-ray emitted from \Cf\ detected by the spectrometer, after the total path $D$.
These events are expected to produce a narrow peak in ToF at 
$\Delta t_{\gamma} = D/c \approx 6.5$~ns.

A possible source of time offset is the fixed delay due to the cable lengths, electronic response and other instrumental 
effects: this is different for each DAQ channel,
but constant throughout the entire data taking. Moreover, it was observed that 
the internal clocks of the FADC boards can be skewed by one or a few 8-ns ticks. 
This offset remains constant throughout each run,
but changes from run to run, due to the re-initialization procedure. Timing corrections were hence modeled using two parameters: 
a channel-specific offset, which is constant across runs, and a run-dependent board offset, common to all channels 
on a given DAQ board. A global least-squares fit was used to determine all offsets by imposing the 
condition that the ToF is equalized at $\Delta t_{\gamma}$ for all 36 \BaF\ --PSci pairs and in all 142~runs. 

\begin{figure} %
\begin{minipage}{\columnwidth}
\centering
 \includegraphics[width=0.85\textwidth]{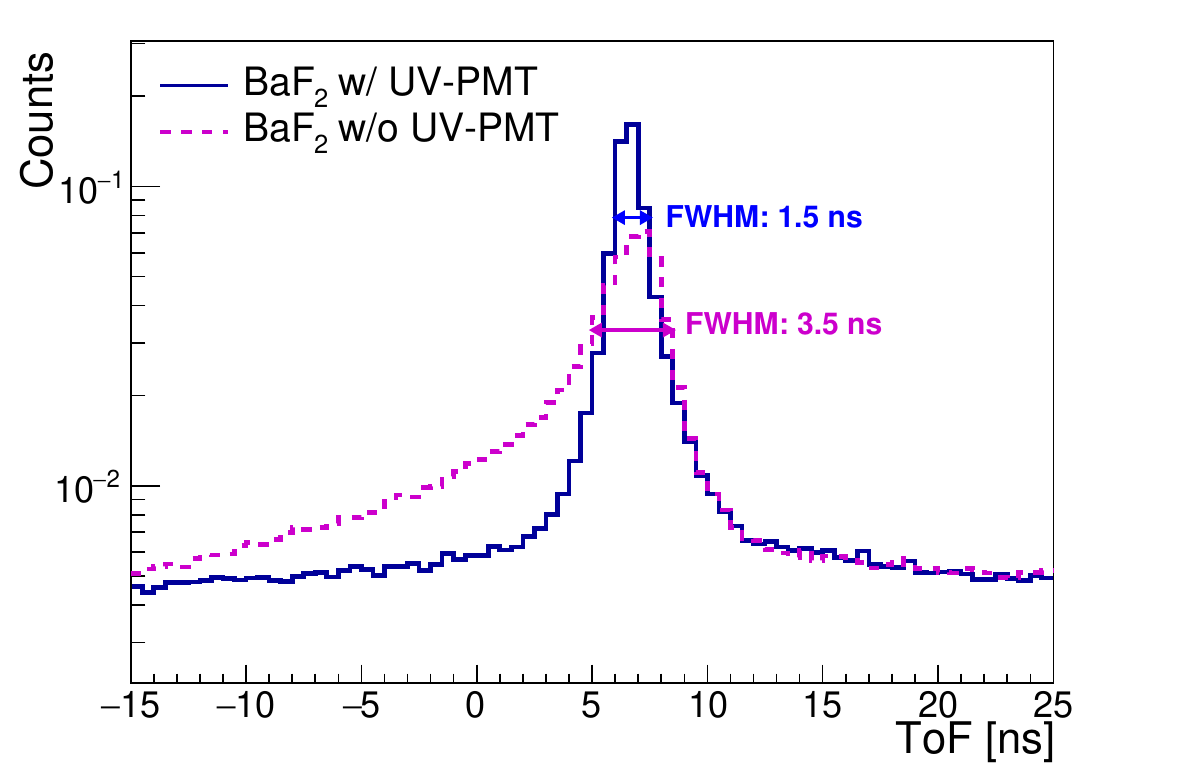}
\end{minipage}
\caption{ToF distributions of $\gamma$-rays from the \Cf\ SF detected in the neutron spectrometer, after the synchronization
procedure described in the text, which sets the peak position at the expected $\Delta t_{\gamma} = 6.5$~ns. The distributions
for BaF0 (blue solid) and BaF1 (dashed violet) are shown separately.}\label{ToF_gamma_flash}
\end{figure}

The ToF distributions for the $\gamma$-ray events between the \BaF\ and the PSci detectors are shown in 
Fig.~\ref{ToF_gamma_flash}, separately for BaF0 and BaF1 and with all corrections applied. The 
weak prompt component observed in the light signal of BaF1 spoils the time accuracy, as it 
produces a significantly broader and non-Gaussian tail in the ToF distribution. The ToF resolution, evaluated as 
the standard deviation of the Gaussian fit of the $\gamma$-ray peak, is 0.6 (1.3)~ns for BaF0 (BaF1), respectively. The full width at half maximum 
(FWHM) of the two distributions is 1.5~ns and 3.5~ns, respectively. The time resolution is sufficient to 
achieve a determination of $K_n$ within better than 5\% in the range 1--10~MeV.
 
\subsubsection{Selection of neutron candidates}
The first step of the data selection is the identification of the tagged neutrons, namely \BaF\ --PSci coincident events in 
which the timing and the PSD are both compatible with a neutron from the \Cf\ source interacting in the PSci. It
is required that: (1) either one of the \BaF\ detectors (or both) is fired and with a reconstructed energy above 100 (200)~keV for BaF0
(BaF1), respectively; 
(2) only one of PSci detectors is fired and with energy above 200~keV, corresponding to about 1~MeV proton 
recoil energy.
In the case that both \BaF\ detectors are fired, the BaF0 is considered for the 
evaluation of the ToF, as it offers superior time resolution, but it is requested that the two times agree within 10~ns.
Selected events are those having a ToF in the range $[40,160]$~ns, which is expected for a \Cf\ neutron 
to cover the 2~m distance between the source and the spectrometer. This ToF cut effectively selects neutrons
with kinetic energy between 0.8 and 12 MeV. Despite the $\gamma$-rays directly coming 
from the source can be easily rejected by ToF, the rate of $\gamma$-induced accidental coincidences within the neutron ToF
range is a factor of 10 larger than the genuine neutron rate. These $\gamma$-rays are 
originated either from the \Cf\ source or from environmental background.
The n/$\gamma$ discrimination capability of the PScis is fundamental to suppress 
the accidental $\gamma$ background: only events in which \fprompt\ is compatible with a neutron are preserved. As 
the performance of the \fprompt\ selection gets worse at lower energy, an energy-dependent
\fprompt\ cut is used. 
The cut is tuned to minimize the contamination of mis-identified $\gamma$ events to less than 1\% of the total selected sample. 
%

\begin{figure} 
\begin{minipage}{\columnwidth}
\centering
 \includegraphics[width=\textwidth]{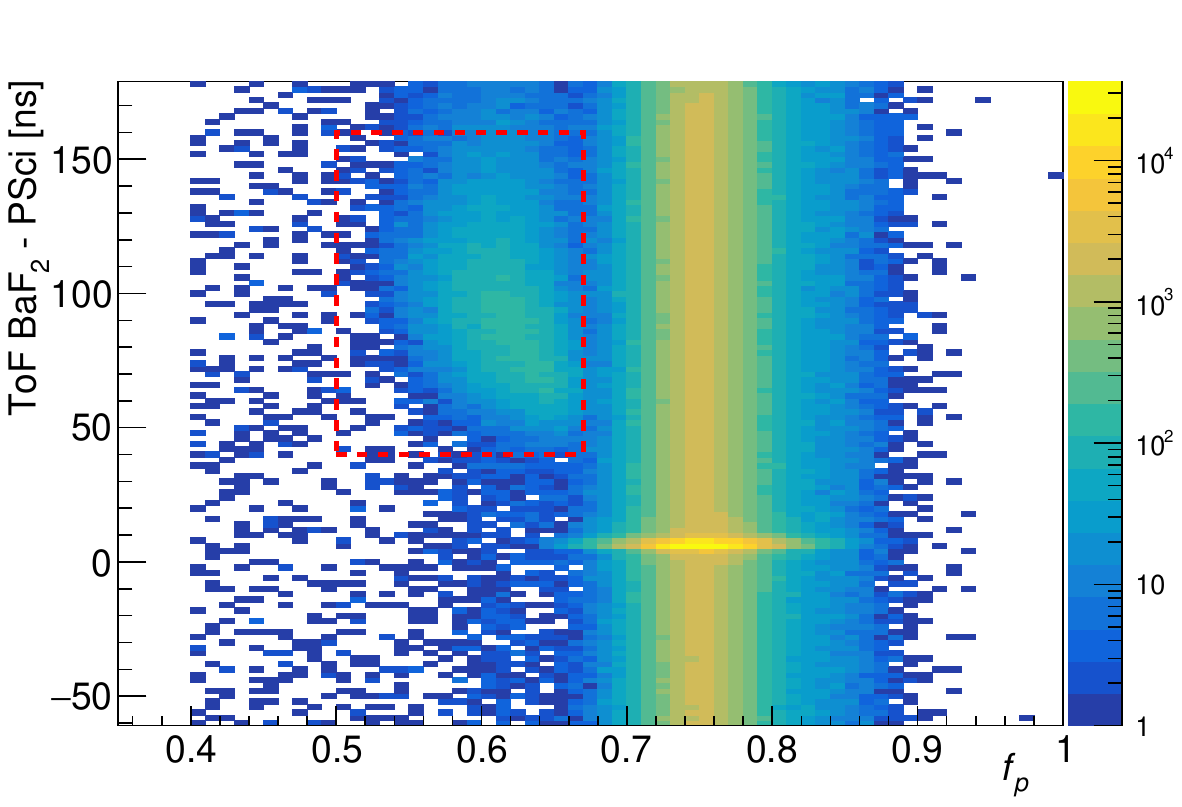}
\end{minipage}
\caption{ToF between \BaF\ and PScis vs.  \fprompt\ of the PSci. The dashed rectangle indicates 
the selection region for tagged neutron events; this is only for visual purposes, as the selection cut in \fprompt\ 
is energy-dependent, see text for more details. The large majority of triggered events comes from SF $\gamma$-rays that reach 
the neutron spectrometer (ToF $\sim 6.5$~ns and \fprompt\ $\sim 0.75$) and $\gamma$-ray accidental coincidences. 
}
\label{ToF_vs_PSD}
\end{figure}

The ToF vs. \fprompt\ scatter plot is shown in Fig.~\ref{ToF_vs_PSD}: the dominant $\gamma$-ray background is the 
band at $f_p \sim 0.75$, while tagged neutrons are the population with ToF between 40 and 160~ns and 
$f_p \lesssim 0.65$. About 50,000 events (0.4\% of the total acquired events) are classified 
as tagged neutrons: for these events the analysis follows up, to check for the presence of a valid signal in the TPC. 

\subsection{Reconstruction of TPC events} \label{subsect:tpcreco}
For the sample of events classified as tagged neutrons, the 28 digitized waveforms from 
the SiPMs of the TPC are analyzed, to search for a possible scintillation and/or ionization signal.  

The time response of the SiPMs to one photoelectron (PE) 
exhibits a prompt and a delayed component, the latter having a time constant of about 600~ns. The response function 
of each individual SiPM is derived from the laser calibration runs and parametrized 
using the analytical model of Ref.~\cite{Agnes:2018hvf}.  
It is numerically deconvolved from the SiPM waveforms in order to produce narrow and quasi-Gaussian PE peaks. The total charge and 
time of the PE peaks are stored on disk for later analysis; due to the 2-ns sampling period of the digitizer, 
the interpolated PE time is also calculated and stored for improved timing. 
The deconvolution procedure was tested in detail and 
verified to preserve the charge information of the original waveform with accuracy better than 99\%.
The single photoelectron (SPE), defined as the charge corresponding to a one PE signal, is derived
from the laser calibrations for each SiPM individually, following the procedure described in Ref.~\cite{Agnes:2021zyq}.
The SPE values are found to be stable within $< 2\%$ throughout the entire data taking period.

The deconvolved waveforms are summed and then scanned by a pulse-finder algorithm based on a moving average technique, to identify 
S1, S2 and possibly echo pulses (S3). Echoes are signals generated by the photoionization of the cathode by the UV 
light of intense S2 signals: one or more electrons are extracted from the cathode, which are eventually drifted to 
the multiplication region and generate a further electroluminescence pulse, delayed 
by $T_{\mathrm{driftmax}}$ with respect to S2.
The pulse finder was tuned by using Monte Carlo (MC) generated waveforms in order to optimize the acceptance 
of weak signals and to minimize false positives. 
MC waveforms were also used to evaluate the pulse finder acceptance of NR signals. For S1, it is $50\%$ at about 15~PE, 
increasing to 100\% for signals above $50$~PE. For S2, the acceptance exceeds $80\%$ at 40~PE (corresponding to approximately 2 electrons) and reaches 100\% above 80~PE.
Single-electron S2 pulses are identifiable with small 
efficiency, about 20\%. 

The time of the (fast) S1 pulses is evaluated by a weighted average 
of the arrival time of the first PE in each SiPM, while S2 and echo times are evaluated by using a constant fraction 
algorithm.

The signal amplitudes are calculated by integrating the total charge of the pulse within a range which is 
dynamically adjusted by the pulse-finder to account for tails, which is typically 10 (30)~\textmu s wide for S1 (S2),
respectively. They are then converted from charge to PE by using the SPE values from the nearest laser calibration.
The S1 light yield measured with the 60-keV $\gamma$-rays from \Am\ is $(7.49 \pm 0.17)$~PE/keV at 
200~V/cm drift field. 
The S2 signal is corrected to account for the presence of impurities, which can
cause the absorption of electrons during their drift path, as described in Ref.~\cite{DarkSide-20k:2023nla}. 
The electron life time was estimated to be larger than 1~ms
from the \Am\ calibration runs: due to the very long life time, 
the correction on the S2 amplitude is within a few percent. The weekly \Am\ calibrations
confirm the time stability of the TPC response, in both S1 and S2, within 2.5\%.
The S2 reconstruction procedure is benchmarked through the detailed MC 
simulation described in Sect.~\ref{sec:MC}.
It is found that the reconstructed S2 is slightly underestimated with respect to the ``MC truth'' value
fed as the MC input, S2/S2$_{true} = 0.99$, because of late PEs 
outside the signal integration window. The bias is corrected a posteriori by scaling the reconstructed S2 
by an efficiency factor of 1.01.
The global accuracy in the reconstruction of the S2 amplitude is better than 5\% rms. The S2 signal is finally
converted from PE to number of electrons, \Nel, by the S2 gain \gtwo, which is discussed in Sect~\ref{subsect:g2}.


The knowledge of the interaction position within the TPC is important for an improved calculation of the neutron 
kinetic energy. The $z$ coordinate is inferred by the drift time \tdrift.
For the TPC events lacking a valid S1 signal (i.e. S2-only), \tdrift\ is evaluated by considering the S2 and \BaF\ signal times: this is an excellent approximation, as the time delay between the \BaF\ and 
the S1 signals, due to the neutron ToF, is a few tens of ns, much shorter than the \textmu s-long drift time of 
electrons within the TPC. The \xy\ position is estimated with a likelihood method that uses the S2 signal pattern detected by the
individual 24 SiPMs of the top tile. The procedure is calibrated using events from environmental background, which are uniformly 
distributed within the TPC, in order to account for the dependence of the S2 amplitude vs. \xy\ originated by non-uniformities 
of light collection and electric field, especially close to the edges. The \xy\ correction for the S2 
amplitude 
is within 4\% in the $4 \times 4$~cm$^2$ central region of the TPC which is considered for the following 
analysis. The accuracy of the \xy\ reconstruction for 
low-S2 events 
within the fiducial volume was evaluated by artificially downsampling the 
S2 signal in the SiPMs from higher energy 
events and validated by using samples of MC generated events. Studies with MC events confirm that the \xy\ 
reconstruction is unbiased and has a resolution of about 5-6~mm rms for low-S2 events ($< 100$~PE, corresponding to 
$N_e \lesssim 5$) and improves to 2.5~mm rms for higher S2. A sub-mm resolution is achieved in $z$, as \tdrift\ 
can be measured with precision significantly better than 1~\textmu s.  

\subsection{Signal event selection} \label{subsect:finalcuts}
A valid TPC signal is found in 2258~events, which is about 5\% of the candidate tagged neutron events.
According to MC, this sample contains a large fraction ($\sim 50\%$) of genuine signals, i.e. events with
a single (n,n') interaction taking place in the TPC. The contamination of accidental
events, in which the TPC signal is uncorrelated with the neutron detected by the \BaF\ and PSci, is less than 1\%, as estimated
by dedicated runs taken with a random trigger.

The sample of signal candidates is further selected by applying a number of cuts based on the TPC event topology.
Firstly, it is requested that there is only one S2 pulse (with a possible echo after $T_{\mathrm{driftmax}} $), which must occur
within the range [5,65]~\textmu s from the \BaF\ time. TPC events with a S2 delay significantly longer than the maximum drift time
 cannot be physically correlated to the tagged neutron. 
A S1 signal in the TPC is allowed, but not required. The expected S1 signal for the events of interest is very small ($\lesssim 10-20$~PE)
and is therefore often missed by the pulse finder. 
For NRs energies close to $\sim 10$~keV, it is however possible that a valid S1 signal is detected for genuine signal events,
albeit with low efficiency. In the case a S1 pulse is also found, it is requested that it is smaller than 100~PE and
that its time is consistent with the neutron ToF, within 15~ns tolerance. This additional time cut allows to reject
inelastic scattering events (n,n'$\gamma$) in the TPC and accidental coincidences. 

An additional cut is placed to discard events in which the SiPMs of the TPC detect in the 5~\textmu s
pre-trigger region an amount of light which is above 0.5~PE/\textmu s, to be compared with the average dark count rate of
about 0.15~PE/\textmu s. The anomalous amount of light is an indicator that the TPC signal could be sitting on the tail
of a previous (and un-trigged) one, which could potentially bias the measurement of the small S2 signal expected for low-energy NRs. 

Finally, as in the previous analysis of Ref.~\cite{Agnes:2021zyq}, a fiducial cut is applied on the reconstructed \xy\ position of the interaction in the TPC: only events within the
inner $4 \times 4$~cm$^2$ region, i.e. at least 1~cm away from the borders, are accepted. The cut removes events located
at the edges of the TPC,
where the increased electric field non-uniformity effects could potentially degrade the accuracy of the S2 measurement.

The topology selection discards about 60\% of the initial candidates. Most of the reduction comes from the fiducial volume and from S2 time 
cuts. 

\subsection{Recoil energy reconstruction} \label{subsect:erec}
For the sample of candidate signals, the recoil energy \Er\ in the TPC is determined event-by-event through Eq.~\ref{eq:recoil_energy}.
The neutron kinetic energy is firstly evaluated using the measured ToF and the total path $D$ traveled
by the neutron. The path $D$ is assumed to be the composition of two straight legs: one from the \Cf\ source to the
interaction point $(x,y,z)$ within the TPC, and one from the TPC
interaction site to the center of the PSci which was fired. The tiny neutron energy loss in the TPC (a few keV out
of $\sim$~MeV) is neglected in the calculation. 
Similarly, the scattering angle $\theta_S$ is evaluated as 
the angle between the two straight legs above, i.e. the angle between the (source-TPC) direction and the (TPC-PSci) direction. 
Only events with reconstructed $E_r$ between 1 and 10~keV are considered
for the subsequent statistical analysis: this is because events outside this range are kinematically incompatible with a single 
(n,n') scattering. 

\begin{figure} 
\begin{minipage}{\columnwidth}
\centering
 \includegraphics[width=0.9\textwidth]{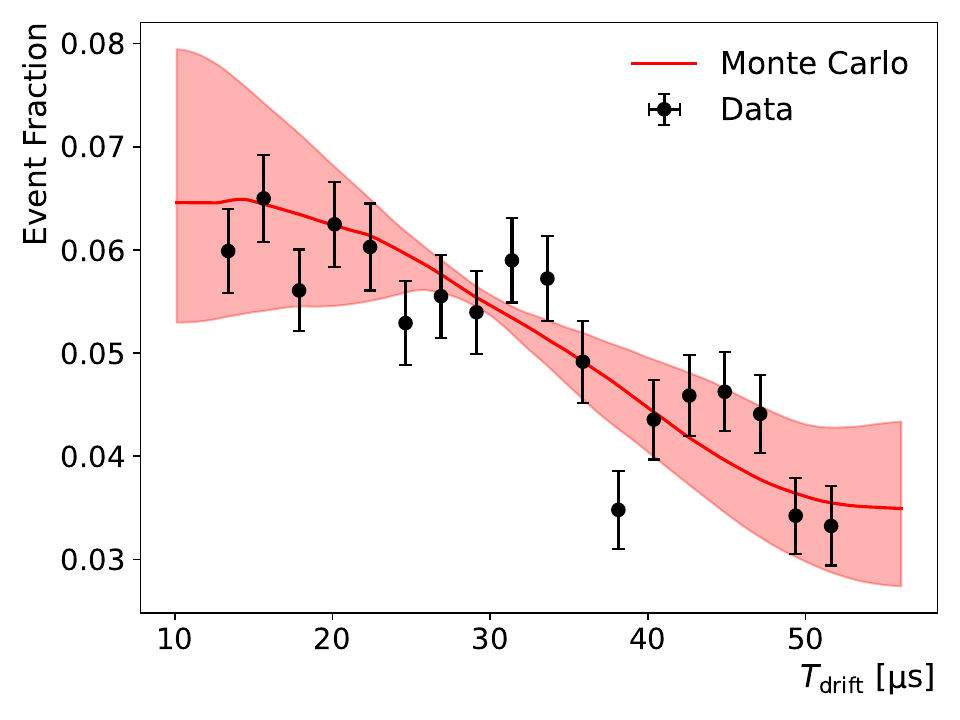}
\end{minipage}
\caption{Drift time distribution from \Am\ calibrations (data points), superimposed with the MC simulation assuming the 
best-fit TPC offset $\Delta z= 0.23$~cm (red line) and the $\pm 1 \sigma$ band (shaded red region).}
\label{fig:am241fit}
\end{figure}

The precise alignment of the setup is a key aspect, as it directly affects 
the evaluation of $E_r$. As discussed in Sect.~\ref{sec:Detectors}, the alignment is at the few-mm level for the source and 
spectrometer, but less binding for the TPC. Being the two PSci arrays deployed symmetrically in the vertical direction, 
the $E_r$ reconstruction is sensitive to a possible mis-alignment to the TPC 
in the $z$ direction\footnote{A 1-cm vertical displacement of the TPC corresponds to a 8\% variation in the reconstructed $E_r$.}. For this 
reason, a possible vertical offset of the TPC with respect to the nominal level is constrained by using \Am\ calibration data. The  
\Am\ source was placed in a small shelf on the external wall of cryostat, thus allowing to measure its $z$ coordinate with respect to 
the center of the spectrometer. The experimental \tdrift\ distribution from the 60-keV $\gamma$-ray 
events from \Am\ is compared against a set of MC simulations, in which vertical displacements 
of the TPC between $-3$ and 3~cm are assumed. The $\Delta \chi^2$ between the experimental and simulated distributions was used to constrain the 
vertical offset of the TPC: the best fit was found at $\Delta z= (0.23 \pm 0.95)$~cm, see Fig.~\ref{fig:am241fit}.
The reconstruction of the recoil energy is hence performed assuming that the TPC is placed 0.23~cm
above the nominal position. A possible displacement of the TPC along the transversal plane \xy\ has a sub-percent effect and it
is neglected here. The effect of the uncertainty of $\Delta z$ on $E_r$ is taken into account as a systematic uncertainty
in Sect.~\ref{sub:systematics} below.

\begin{figure*} 
\centering
\includegraphics[width=0.33\textwidth]{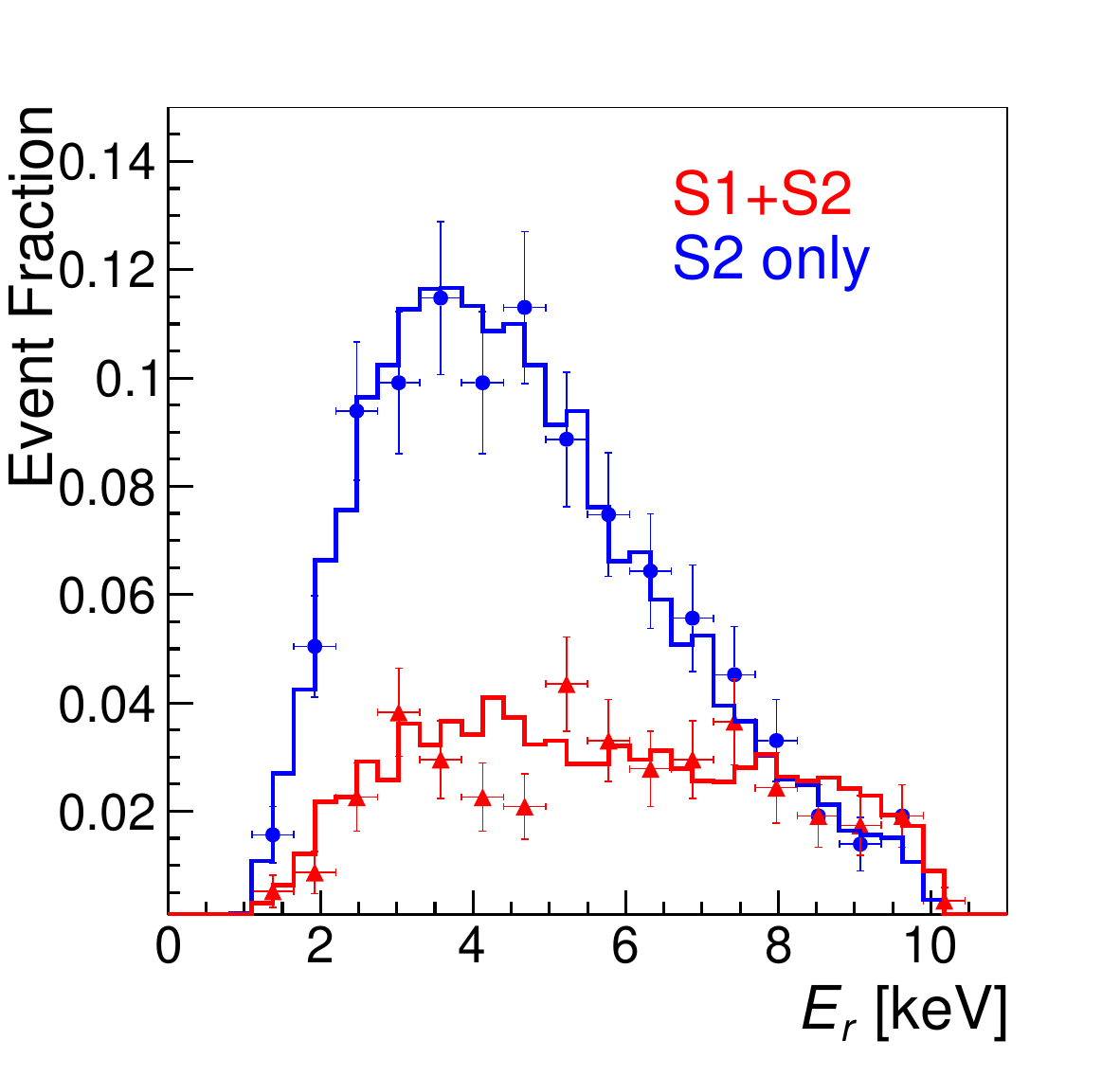}
    \hfill
    \includegraphics[width=0.33\textwidth]{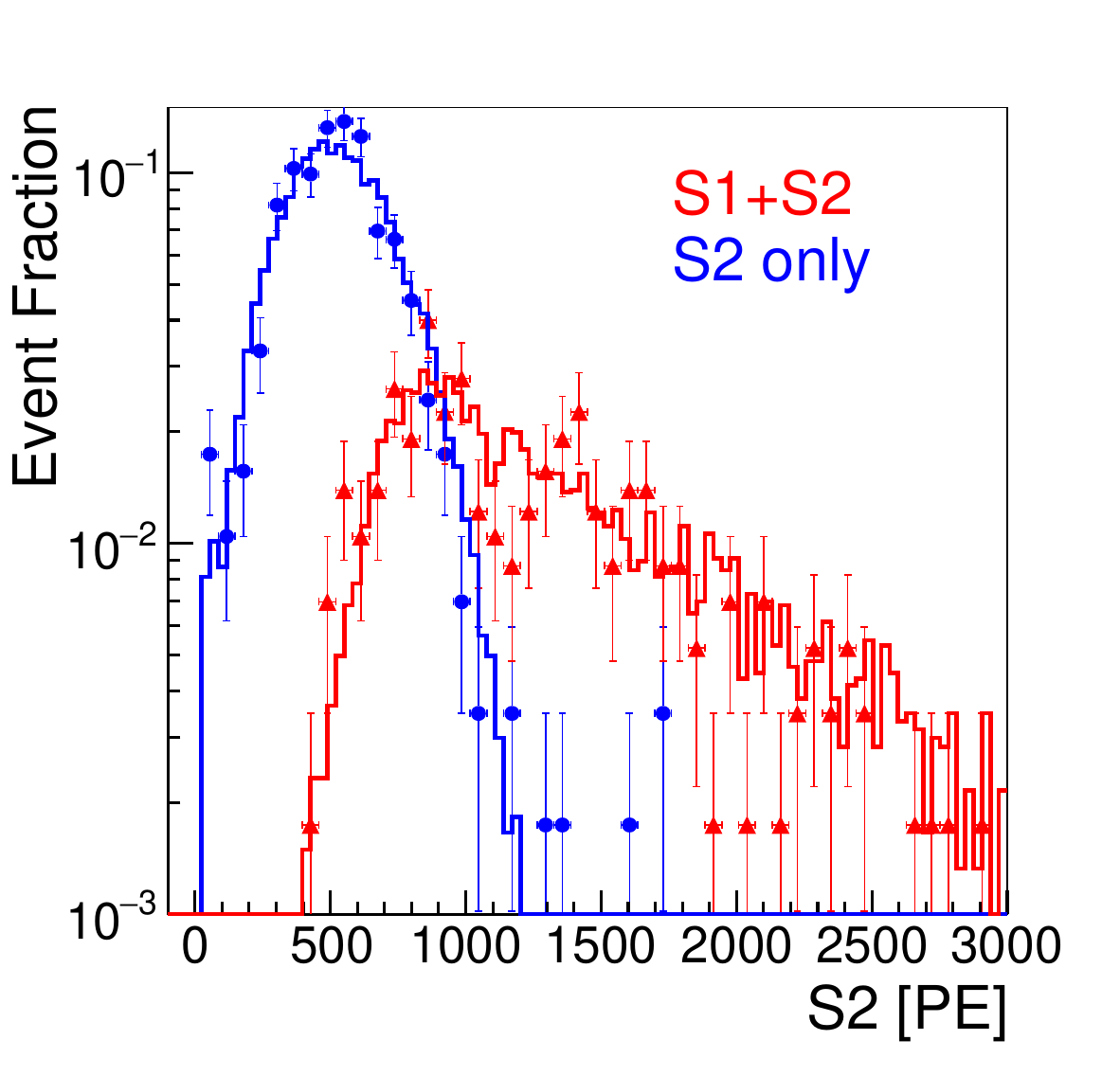}
    \hfill
    \includegraphics[width=0.33\textwidth]{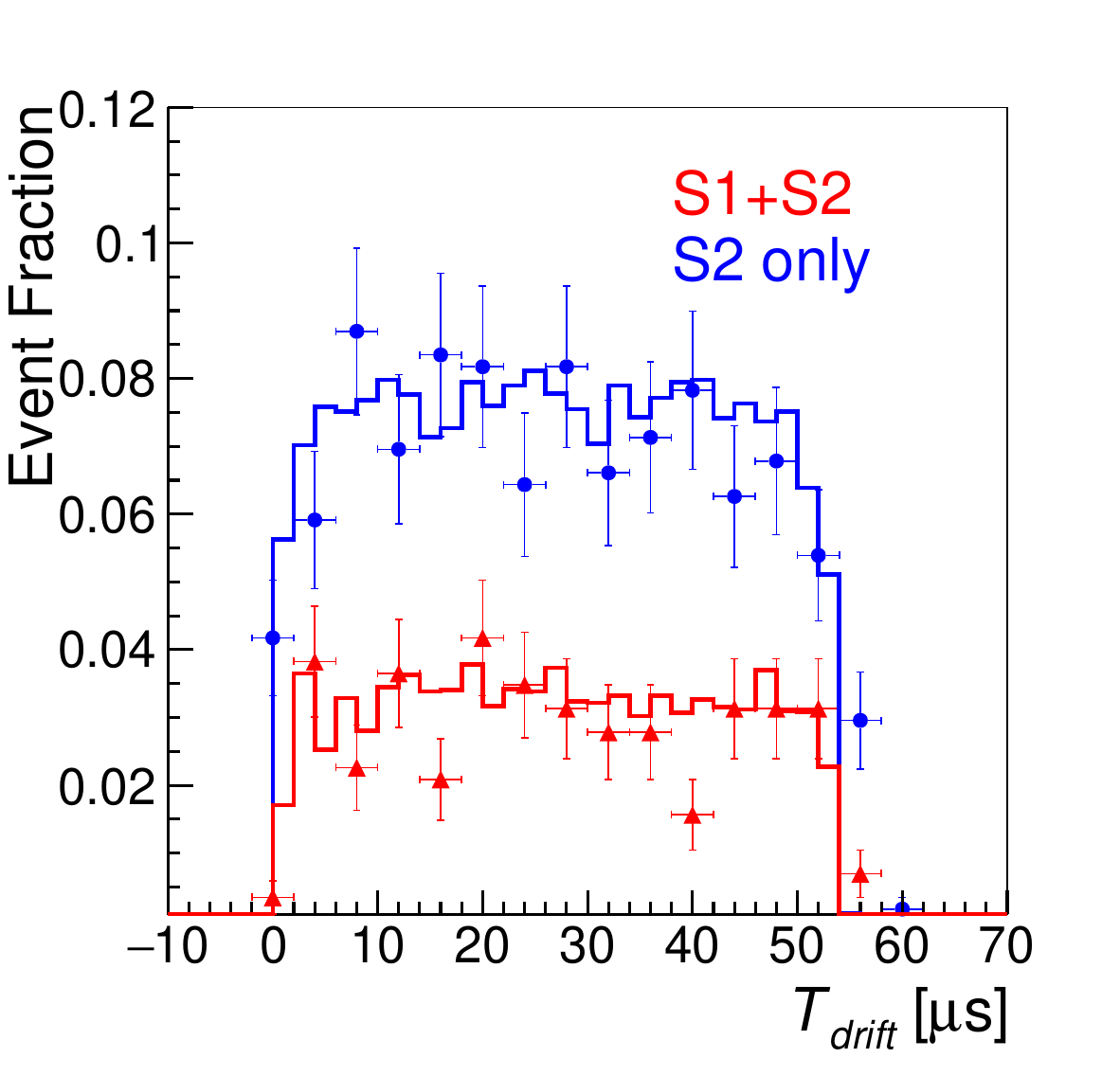}
\caption{Distribution of the recoil energy \Er\ (left panel), S2 (middle panel) and \tdrift\ (right panel) for the final sample
of candidate signal events. The S1+S2 (red triangles) and the S2-only (blue circles) populations are shown separately and
superimposed with the corresponding MC simulation (solid histograms).}
\label{sample_diagnostic}
\end{figure*}

\begin{figure} 
\begin{minipage}{\columnwidth}
\centering
 \includegraphics[width=0.99\textwidth]{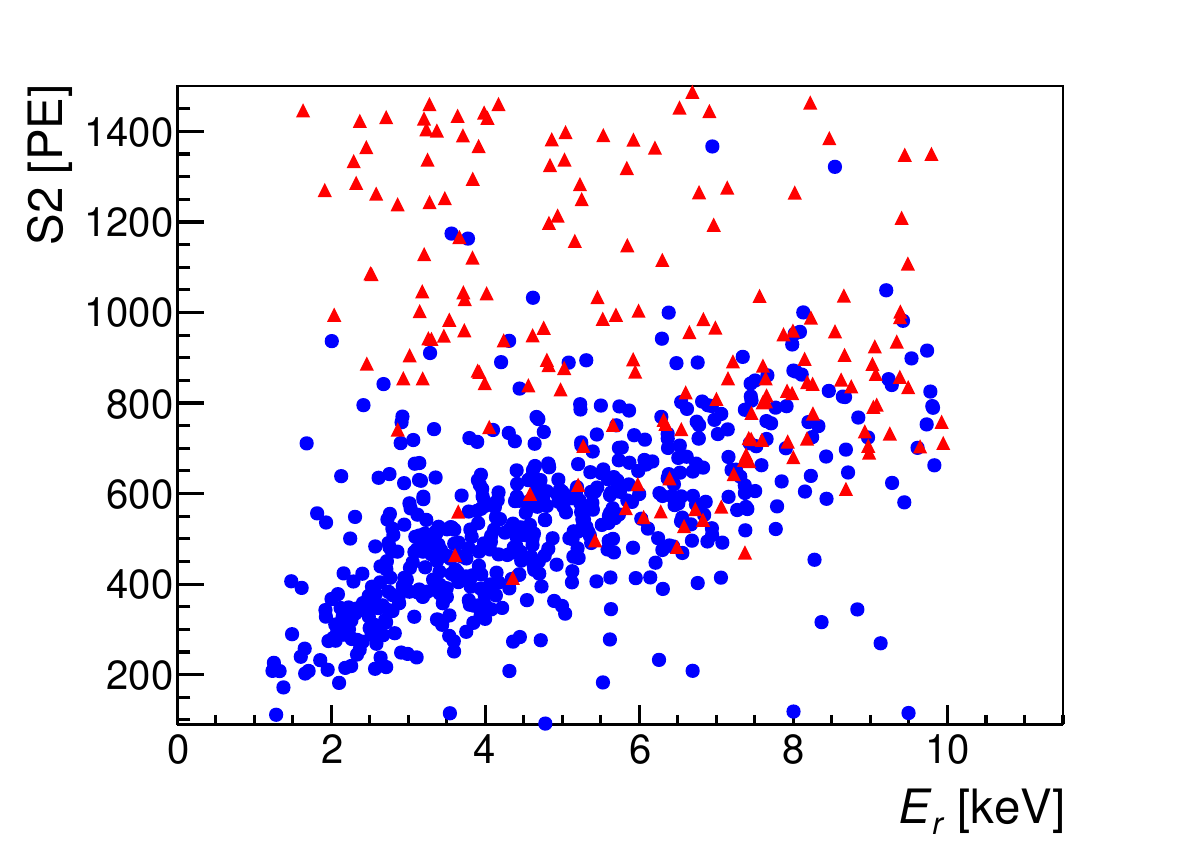}
\end{minipage}
\caption{Scatter plot of S2 vs. \Er\ for the final sample of candidate signal events. S1+S2 and S2-only events are represented separately 
as red triangles and blue circles, respectively.}
\label{s2er_scatter}
\end{figure}

The final sample considered for the statistical analysis contains 806 events with reconstructed energy 
between 1 and 10~keV: 575 S2-only (71\%) and 231 S1+S2. The fraction of S2-only events is well compatible 
with the expectations from the MC simulation, 70\%. The scatter plot of S2 vs. \Er\ for the final sample is 
shown in Fig.~\ref{s2er_scatter}. The distributions of $E_r$, S2 and \tdrift\ are displayed in Fig.~\ref{sample_diagnostic}, separately 
for the S1+S2 and S2-only events. 
According to the MC simulation, approximately 2/3 of the S2-only population consists of signal events from single neutron scattering. In contrast,
the majority (about 80\%) of S1+S2 events originate from background due to 
multiple neutron scattering, in which the kinematical correlation is lost. These events typically exhibit a significantly
higher S2 than those in the S2-only population. The \tdrift\ distribution is flat, as expected from 
a uniform neutron illumination throughout the entire sensitive volume of the TPC. The \Er\ distribution peaks between 3 and 5~keV and 
extends down to $\sim$1~keV, thus confirming that ReD could meet its design goal.

\section{Monte Carlo simulations} \label{sec:MC}
A detailed Monte Carlo simulation based on 
\textsc{Geant4}~\cite{Agostinelli:2003fg,Allison:2006cd,Allison:2016lfl} was 
developed to validate the experimental layout and to model the detector response. 
The simulation initially produces the 
energy deposits in all active volumes, following the SF of \Cf. The 
prompt neutron spectrum is modeled according to the data-driven Mannhart 
distribution~\cite{Mannhart1987}.
The prompt $\gamma$ component from 
the fission 
is not known with comparable accuracy: $\gamma$-ray multiplicity and energy 
distributions were modeled from Refs.~\cite{brunsonthesis,wagemans_1991,Valentine:2001}.
As the simulation is extremely CPU-intensive due to the thick neutron shielding, a biasing 
approach is employed, which favors neutrons entering the collimator cone.

The energy deposits within the TPC are then used to simulate the number of 
scintillation photons $N_\gamma$ and of ionization electrons $N_e$ surviving 
recombination using a model tuned on DarkSide-50 data~\cite{PhysRevD.104.082005}, which  
also accounts for fluctuations. 
The expected light signals in the detector are finally evaluated as 
S1 = $ g_1 N_\gamma$ and  S2 = $ g_2 N_e$, with parameters 
$g_1 = 0.146$~PE/photon 
and $g_2 = 18.6$~PE/e- evaluated from calibrations (see Sect.~\ref{subsect:g2}).
The simulation also produces the time list of all individual PEs within S1 
and S2. A realistic TPC synthetic waveform is finally produced for each event 
by convolving the PE time distribution above with the SiPM time response 
function from Ref.~\cite{Agnes:2018hvf} and by summing a baseline with 
the same standard deviation as experimental data. Realistic waveforms 
for the \BaF\ and PSci detectors are produced by a similar procedure. The 
response functions are tuned on the experimental data in order to produce 
simulated signals that are as realistic as possible.  Particular care was 
devoted to reproducing the individual time responses of the two \BaF\ detectors, 
which are remarkably different, see Fig.~\ref{ToF_gamma_flash}.

The synthetic signals produced by the simulation undergo the same 
reconstruction chain and analysis flow as the real data.
The simulation is benchmarked by successfully reproducing the shape of a large variety of experimental 
distributions, as the energy spectra of each \BaF\ detector; the PSci energy distribution in multiple
ranges of ToF; and the ToF for tagged neutron events.
The SF \Cf\ activity required to match the absolute 
normalization of the simulated and experimental neutron spectra in the \BaF\ is 19.5~kBq,
which is in reasonable agreement with nominal value provided by the manufacturer, 
25~kBq ($\pm 15\%$).
The simulation of the TPC events is verified for both electron and neutron events 
using a dedicated \Cf\ data set acquired with a coincidence trigger condition between 
the \BaF\ and the TPC. 

Following validation, the simulation is used to evaluate efficiencies, resolutions and other 
systematic uncertainties. Specifically on the TPC reconstruction, simulated data samples 
are employed to evaluate the pulse finder efficiency for S1 and S2 signals, as well as to
verify the resolution and the possible bias of the reconstruction algorithms.
The accuracy of the reconstruction of the number of electrons $N_e$ is much better than 1\% for
$N_e > 10$. The $N_e$ resolution ranges between 12\% for $N_e = 10$ to about 7\% for $N_e > 40$, 
accounting globally for the S2 signal corrections (amplitude and \xy), and for the $N_e$
Poisson fluctuations.

The energy reconstruction procedure of Sect.~\ref{subsect:erec} is confirmed to be accurate for 
genuine single-scattering signal events, as $\langle E_r/E_r^{true} \rangle = (0.997 \pm 0.001)$. 
The resolution $\delta E_r$ ranges from 9\% at $\sim 2$~keV to 6\% above 8~keV. This is mostly 
driven by the size of the PScis, as the lack of information about the actual neutron interaction 
position within the detector affects the determination of both $K_n$ and $\theta_S$. 
Sub-leading contributions to $\delta E_r$ come from the ToF resolution and from the accuracy of the 3D reconstruction 
within the TPC. 


\section{Data analysis} \label{sec:StatisticalAnalysis}
\subsection{Determination of the ionization gain \gtwo} \label{subsect:g2}
The S2 signal is converted into the number of ionization electrons \Nel\ by dividing it by the ionization gain \gtwo. This detector-specific parameter represents the average S2 response produced by a single electron extracted into the gas phase. 
The value of \gtwo\ is determined from \Am\ calibration data.
Firstly, TPC events are selected from \Am\ calibration data 
featuring (1) one S1 and one S2 signal; (2) S1 amplitude compatible with the 60-keV $\gamma$-ray; (3) reconstructed 
\xy\ position within the central $4 \times 4$~cm$^2$ square; (4) reconstructed $z$ position within the central half of 
the active detector. The S2 distribution of these events was fitted with a Gaussian to determine the mean S2 value 
$\langle \text{S2}^\textrm{Am} \rangle = (13380 \pm 120_{\textrm{stat}})$~PE. 
The expected number of electrons $\langle N_{e}^\textrm{Am} \rangle$ is estimated by using the MC simulation of 
Sect.~\ref{sec:MC} as
\begin{equation}
\langle N_{e}^\textrm{Am} \rangle = \sum_i \epsilon_i Q_{y,ER}(\epsilon_i) \label{g2sum}
\end{equation}
where $\epsilon_i$ is the energy released in each interaction site within the TPC and $Q_{y,ER}(\epsilon_i)$ is the 
energy-dependent ionization yield for electron recoils, 
modeled according to the semi-empirical DarkSide-50 parametrization of Ref.~\cite{PhysRevD.104.082005}. 
The summation of Eq.~\ref{g2sum} is restricted to the events 
which meet the selection criteria above. The average number of electrons for \Am\ events 
is $\langle N_{e}^\textrm{Am} \rangle = (721.13 \pm 0.21_{\textrm{stat}})$. 
The parameter \gtwo\ is taken as the ratio between $\langle \text{S2}^\textrm{Am} \rangle$ and 
$\langle N_{e}^\textrm{Am} \rangle$, namely 
\begin{math}
g_2 = (18.56 \pm 0.17_{\textrm{stat}} \pm 0.69_{\textrm{syst}}) \ \textrm{PE/e-}, 
\end{math}
or
\begin{equation} 
g_2 = (18.56 \pm 0.71) \ \textrm{PE/e-}, \label{eq:gtwo}
\end{equation}
with statistical and systematic uncertainties summed in quadrature. The 
systematic uncertainty accounts for: uncertainty on the $Q_{y,ER}(E)$ profile from 
Ref.~\cite{PhysRevD.104.082005}; variation in the binning and range of the S2 fit; variation in the S1 range used to 
select the \Am\ events; uncertainty on the SPE; \xy\ correction of the S2 response. 

The estimate of \gtwo\ is double-checked independently by the analysis of echo signals S3, that are 
delayed by $T_{\textrm{driftmax}}$ with respect to S2 time, $t_{S2}$. S3 signals are typically made by 
one or a-few electrons and they are too small to be efficiently identified by the pulse finder:  
their amplitude is evaluated from the charge integrated in a fixed 20~\textmu s window starting 
from ($t_{S2} + T_\textrm{driftmax}$). The S2 signals,
however, present a long tail which extends for several tens of \textmu s after $t_{S2}$ and which 
is attributed to delayed photon emission. This delayed photon tail is of comparable size as the echo signal, 
so its proper subtraction requires a full modeling of the time development of the entire 
S2+S3 signal. An empirical model has been developed for this purpose, which provides a best fit for \gtwo\ 
of $(18.1 \pm 0.5_\textrm{stat})$~PE/e-, in good agreement with the estimate based on the \Am\ calibration. 
As the systematic uncertainties of the model cannot be easily evaluated, this estimate is regarded as a 
corroboration of the value of Eq.~\ref{eq:gtwo} but it is not used in the following.

\subsection{Statistical analysis and systematic uncertainties} \label{sub:systematics}
\begin{figure} %
\begin{minipage}{\columnwidth}
\centering \includegraphics[width= 0.7\textwidth]{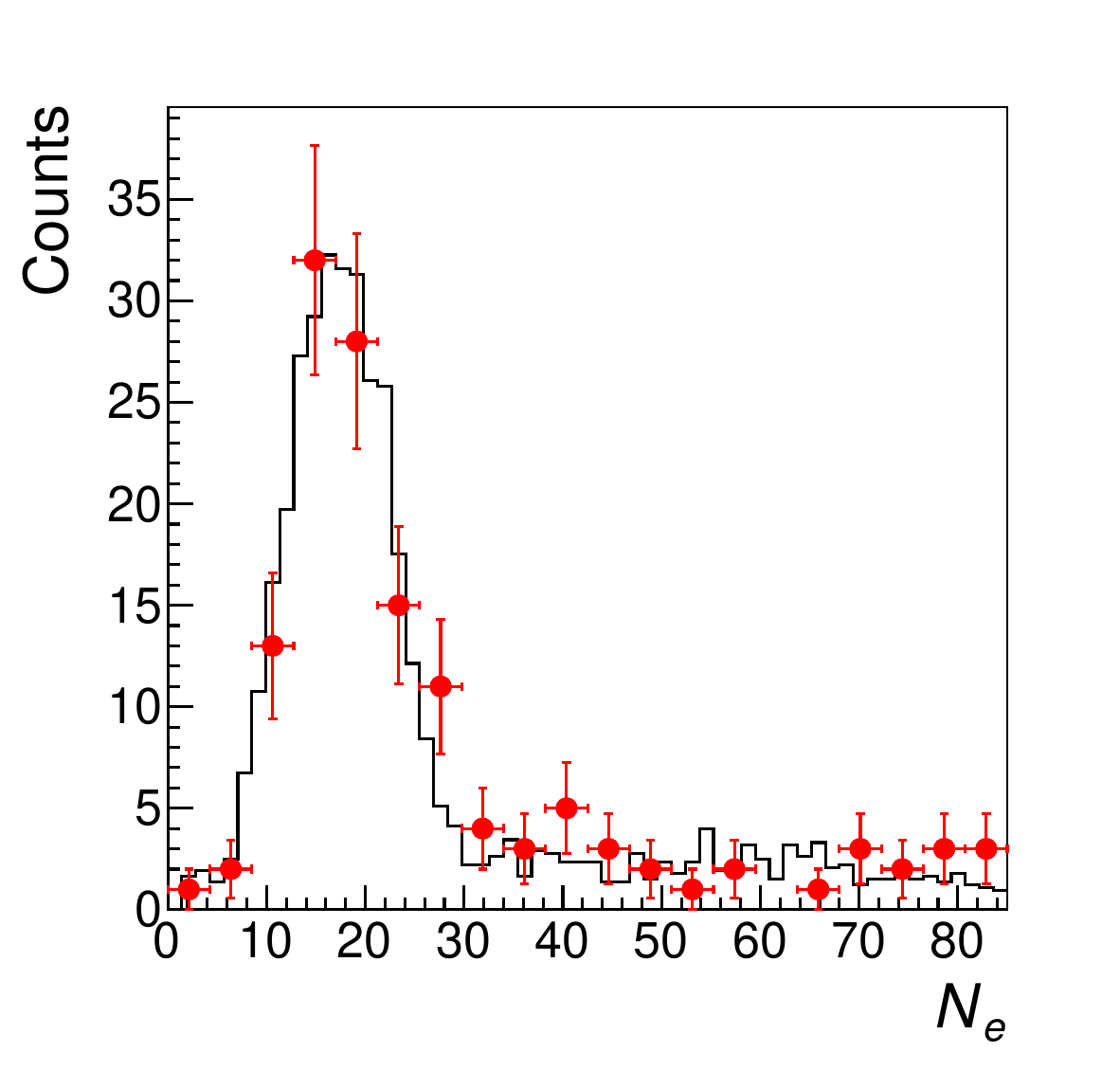}
\end{minipage}
\caption{Experimental \Nel\ distribution for NRs in $1 < E_r < 3$~keV (red circles), corresponding to the first interval 
of Table~\ref{tab:results}, superimposed with the corresponding MC simulation (black histogram). The \Nel\ from the MC, 
which depends on the \Qy\ model fed in input, is rescaled by about 3\% in order to match the position of the 
experimental peak.}
\label{fig:Neslice}
\end{figure}

The data sample presented in Fig.~\ref{s2er_scatter} is partitioned into five sub-samples based on the
reconstructed NR energy \Er, as listed in Table~\ref{tab:results}. Within each energy interval, the number of electrons
\begin{math}
N_e = S2/g_2
\end{math}
is calculated for all events to produce a \Nel\ distribution as shown for example in Fig.~\ref{fig:Neslice}.
This distribution is fitted using an empirical model composed of a Gaussian for the signal and a constant
component for the background. Due to the limited statistics in each interval, an unbinned maximum likelihood
fit is performed. The best-fit value of the Gaussian peak, $N_e^{fit}$, serves as the estimator
for the mean electron count in the interval.
Since the distribution of event energies within each sub-sample is not uniform, the mean
energy \Erbar\ is used instead of the interval's central value. The ionization yield \Qy\ at \Erbar\
is finally evaluated as the ratio 
$Q_y(\bar{E}_r) = N_e^{fit}/\bar{E}_r$. 

\begin{figure} %
\begin{minipage}{\columnwidth}
\centering \includegraphics[width= \textwidth]{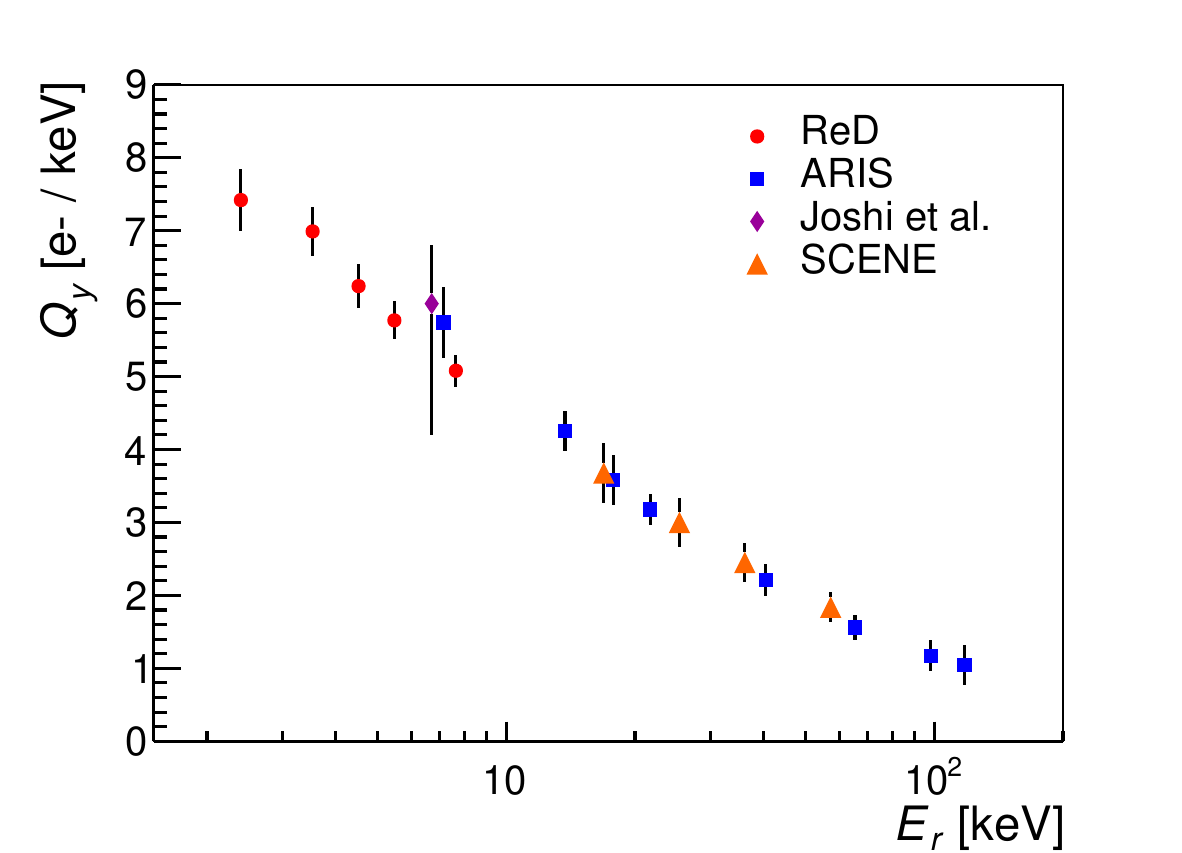}
\end{minipage}
\caption{Ionization yield for NRs obtained from this work between 2 and 10~keV, with
  statistical and systematic uncertainties combined. The data points are overlaid with
  the literature results up to 120~keV by Joshi et al.~\cite{Joshi:2014fna}, ARIS~\cite{Agnes:2018mvl} 
  and SCENE~\cite{Cao:2015ks}.
  The data point at 6.7~keV from Joshi et al. was rescaled according to the discussion presented in
  Ref.~\cite{PhysRevD.104.082005}.}
\label{fig:final_comparison}
\end{figure}

The results, summarized in Table~\ref{tab:results}, are also shown in Fig.~\ref{fig:final_comparison},
overlaid with previous experimental data below 120~keV~\cite{Joshi:2014fna,Agnes:2018mvl,Cao:2015ks}.
Following the discussion of Ref.~\cite{PhysRevD.104.082005}, the experimental
measurement by Joshi et al.~\cite{Joshi:2014fna} at 6.7~keV was rescaled to $6.0^{+0.8}_{-1.8}$ e-/keV, using
the 2.82~keV line from the $^{37}$Ar as a cross-calibration point with the DarkSide-50 data.


%
%

\begin{figure} %
\begin{minipage}{\columnwidth}
\centering
 \includegraphics[width= \textwidth]{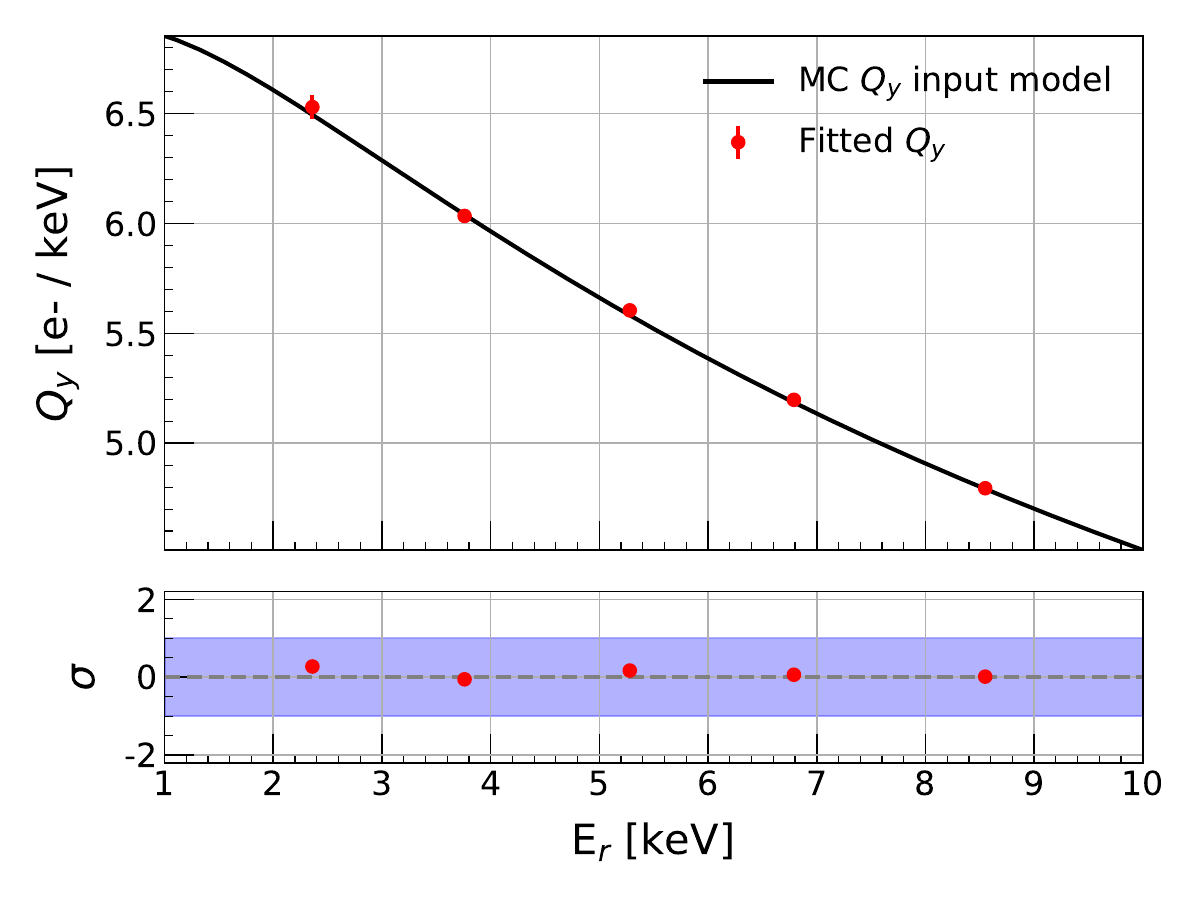}
\end{minipage}
\caption{Validation of the analysis procedure using the sample of MC events. Upper panel: "MC truth" \Qy\ model based on DarkSide-50 data~\cite{PhysRevD.104.082005} used as input for the MC simulation (solid line), superimposed with the fitted \Qy\ and \Erbar\ from the analysis procedure (red markers). The pull terms of the points with respect to the model are displayed in the lower panel, with 
the $\pm 1 \sigma$ band shaded in violet.}\label{fig:mc_bias}
\end{figure}

The validity of the entire analysis chain is verified by using the MC data sample. The data set is analyzed
according to the same 
procedure described above, to evaluate \Erbar, $N_e^{fit}$ and 
\Qy\ in each energy range. The results are in very good agreement with the MC truth, i.e. with the $Q_y(E)$ model 
which was fed in input to the MC simulation, as shown in Fig.~\ref{fig:mc_bias}.  
The analysis reproduces the input $Q_y(E)$ model, calculated at \Erbar, with a relative
difference $< 0.5\%$ and well within one standard deviation.
This study demonstrates that the analysis remains unbiased when both S1+S2 and S2-only populations are included. However,
if just S2-only events are considered, a bias in \Qy\ relative to the MC truth emerges for \Er\ above 6~keV.
%

The uncertainty on \Qy\ is evaluated by propagating the individual uncertainties in $N_e^{fit}$ and \Erbar\ through
\begin{equation}
\left(\frac{\delta Q_y}{Q_y}\right)^2 = \left(\frac{\delta \bar{E}_r}{\bar{E}_r}\right)^2 + \left(\frac{\delta N_e^{fit}}{N_e^{fit}}\right)^2.
 \label{eq:errqy}  \end{equation}
The statistical uncertainty on \Erbar\ ranges between 0.02 and 0.07~keV, depending on the 
width and population of each energy interval.
The main systematic uncertainty on \Erbar\ arises from geometrical misalignment and TPC offset $\Delta z$.
It is estimated to be between 0.04 and 0.06~keV by repeating 
the calculation of \Erbar\ in each energy interval with $\Delta z \pm 1 \sigma$. 
The combined impact on \Erbar\ of all other systematic effects, including the source-TPC and
TPC-spectrometer distances, is below 0.5\%.  
The summation in quadrature of the statistical and systematic terms gives a total uncertainty
$\delta \bar{E}_r / {\bar{E}_r}$ of 2.5\% for the lowest energy interval and about 1.5\% for higher \Er. 

The statistical uncertainty on $N_e^{fit}$ from the likelihood fit is approximately 
0.7~electrons, corresponding to $\sim 4\%$ at the lowest energy interval. This estimate inherently includes
the effects of S2 resolution, \xy\ corrections and Poisson fluctuations on \Nel.
In fact, resolution factors cause a broadening 
of the \Nel\ distribution,
which is intrinsically accounted by the fit to estimate the uncertainty on $N_e^{fit}$. 
The analysis performed with the MC data sample shows that the global effect of all these terms, as 
well as of the shape of the function used to model the background events in the fit, is at the level 
of $<  0.5\%$ and hence negligible with respect to the statistical uncertainties. 
The leading contribution to the systematic uncertainty on $N_e^{fit}$ comes from the 
uncertainty on \gtwo, which produces a coherent shift of all \Nel\ data points: this is 
$\delta g_2/g_2 = 3.8 \%$ and propagates linearly to $N_e^{fit}$. The total uncertainty on $N_e^{fit}$ is thus
between 4 and 5\%. 

Summing all the contributions in quadrature, the final uncertainty on \Qy\ lies between 4.3\% and 5.7\%, primarily driven by the systematic uncertainty on \gtwo. Statistical uncertainties are of comparable magnitude
in the lowest energy intervals only.

\begin{table*}
\centering
    \caption{Final results coming from the unbinned likelihood fit in the five energy intervals listed in 
the first column, each containing $N$ events (column 2). Statistical and systematic uncertainties are quoted 
separately for the mean energy \Erbar\ and for the number of electrons $N_e^{fit}$ in columns 3 and 4, 
respectively. The final values of \Qy\ vs. energy are listed in the last two columns, with statistical 
and systematic uncertainties quoted separately or summed in quadrature, respectively.}    \label{tab:results}
    \begin{tabular}{c c | c|c | c c}
    \toprule
 \thead{\Er\ range} & \thead{$N$} & \thead{\Erbar}  & \thead{$N_e^{fit}$} & \multicolumn{2}{c}{\thead{\Qy}} \\
  (keV) & & (keV)  &  & \multicolumn{2}{c}{(e-/keV)}   \\
 \hline
 $[1.0,3.0]$ & 147  & $2.40 \pm 0.04 \pm 0.04$ & $17.8 \pm 0.6 \pm 0.7$ & $7.42 \pm 0.27 \pm 0.32$ & $7.42 \pm 0.42$ \\
 $[3.0,4.0]$ & 157  & $3.53 \pm 0.02 \pm 0.04$ & $24.7 \pm 0.7 \pm 0.9$ & $6.99 \pm 0.20 \pm 0.27$ & $6.99 \pm 0.34$ \\
 $[4.0,5.0]$ & 125  & $4.52 \pm 0.03 \pm 0.05$ & $28.2 \pm 0.7 \pm 1.1$ & $6.24 \pm 0.16 \pm 0.25$ & $6.24 \pm 0.30$ \\
 $[5.0,6.0]$ & 130  & $5.48 \pm 0.02 \pm 0.06$ & $31.6 \pm 0.7 \pm 1.2$ & $5.77 \pm 0.13 \pm 0.23$ & $5.77 \pm 0.26$ \\
 $[6.0,10.0]$ & 247  & $7.63 \pm 0.07 \pm 0.07$ & $38.8 \pm 0.7 \pm 1.5$ & $5.08 \pm 0.10 \pm 0.20$ & $5.08 \pm 0.22$ \\
\hline
 \bottomrule
    \end{tabular}
\end{table*}

\section{Discussion and future perspectives} \label{sec:discussion}
This work presents the direct measurement of the ionization yield in argon for nuclear recoils between 2 and 10~keV,
as summarized in Table~\ref{tab:results}. 
The region below 6.7~keV is probed for the first time using a direct and model-independent approach.
As shown in Fig.~\ref{fig:final_comparison}, the results from this analysis are in good agreement
with previous measurements above 6.7~keV reported by Joshi et al.~\cite{Joshi:2014fna},
ARIS~\cite{Agnes:2018mvl}, and SCENE~\cite{Cao:2015ks}, supporting a coherent global picture of the
ionization response. The ReD data indicate an increase of the ionization yield at lower recoil energies. 
This trend is qualitatively consistent with the nuclear stopping power model by Lenz and
Jensen~\cite{Lenz1932,Jensen1932}, but in contrast with the model by Ziegler~\cite{Ziegler:2010bzy}, which
predicts a nearly constant behavior of \Qy\ below 5~keV~\cite{PhysRevD.104.082005}. A quantitative
comparison of stopping power models is beyond the scope of this paper and is being addressed in
a dedicated companion publication~\cite{newpaper}.

Beyond the \Qy\ measurements, the data enable new insights into ionization fluctuations for NRs in argon,
which remain poorly characterized in this energy region. These fluctuations are of particular interest for argon-based
Dark Matter experiments, where two limiting cases -- no fluctuations and pure binomial fluctuations -- are
commonly assumed to assess sensitivity to low-mass
WIMPs~\cite{PhysRevD.107.063001,DarkSide-20k:2024yfq,GlobalArgonDarkMatter:2022ppc}. An analysis aimed at
constraining the ionization fluctuation model is in progress and will be the subject of a future publication.

To extend the recoil energy coverage down to 0.5~keV, a new data-taking campaign with \Cf\ is planned for 2026 at
INFN Laboratori Nazionali del Sud (LNS) under the ReD+ initiative. This next-generation effort builds
on the ReD experimental framework and on the same conceptual layout, incorporating upgraded components and
improvements based on the experience of ReD~\cite{lidine2024}.
As a subsequent step, an experimental run targeting NRs down to 0.2~keV will be
conducted using 2.4~MeV neutrons generated via the d(d,$^3$He)n reaction from a commercial Deuterium-Deuterium
(DD) generator. Compared to the \Cf\ source, the DD generator offers a quasi-monoenergetic neutron spectrum,
significantly reduced x/$\gamma$-ray emission, and a neutron fluence up to 10$^7$~n/s. Furthermore,
the detection of the accompanying $^3$He nucleus opens the possibility for event-by-event neutron tagging,
providing substantial background suppression. The generator is currently being commissioned at
the University of S\~{a}o Paulo and will be transferred to LNS following the \Cf\ campaign.

\begin{acknowledgements}

The Authors express their gratitude to the entire technical staff of the INFN Division of Catania.\\
This work has been supported by the PRIN2022 grant 2022JCYC9E, call for tender No. 104 published on 2.2.2022 
of the Italian Ministry of University and Research (MUR) under the National Recovery and Resilience Plan (NRRP), Mission 4, Component 2, Investment 1.1, funded by the European Union – NextGenerationEU, CUP I53D23000690006.  \\
This work is supported by the NCN, Poland (2021/42/E/ST2/00331), the EU’s Horizon 2020 (No 952480, DarkWave project), IRAP AstroCeNT
(Grant No. MAB/2018/7) funded by FNP from ERDF, and the S\~{a}o Paulo Research Foundation (FAPESP) (Grant No. 2021/11489-7). The authors also acknowledge the support of the French Agence Nationale de la Recherche (ANR), under grants ANR-22- CE31-0021 (project X-ArT) and ANR-23-CE31-0015 (project FIDAR), and of IN2P3–COPIN (No. 20-152). 
This work is also supported by the National Key Research and Development Project of China, Grant No. 2022YFA1602001.\\
For the purpose of open access, the authors have applied a Creative
Commons Attribution (CC BY) public copyright license to any Author
Accepted Manuscript version arising from this submission.

\end{acknowledgements}

\bibliographystyle{spphys}       
\bibliography{cfpaper}   



\end{document}